# Nonhuman Primate Brain Tissue Segmentation Using a Transfer Learning Approach

Zhen Lin[1], Hongyu Yuan[2], Richard Barcus[2], Qing Lyu[1,2], Sucheta Chakravarty[2], Megan E. Lipford[2], Carol A. Shively[3], Suzanne Craft[4], Mohammad Kawas[2], Jeongchul Kim[2], Christopher T. Whitlow[1,2]

[1] Department of Biomedical Engineering, Wake Forest University School of Medicine
[2] Department of Radiology, Wake Forest University School of Medicine
[3] Department of Pathology, Wake Forest University School of Medicine
[4] Department of Gerontology and Geriatric Medicine, Wake Forest University School of Medicine

**Abstract.** Non-human primates (NHPs) serve as critical models for understanding human brain function and neurological disorders due to their close evolutionary relationship with humans. Accurate brain tissue segmentation in NHPs is critical for understanding neurological disorders, but challenging due to the scarcity of annotated NHP brain MRI datasets, the small size of the NHP brain, the limited resolution of available imaging data and the anatomical differences between human and NHP brains. To address these challenges, we propose a novel approach utilizing STU-Net with transfer learning to leverage knowledge transferred from human brain MRI data to enhance segmentation accuracy in the NHP brain MRI, particularly when training data is limited. Specifically, we first train our STU-Net model on the Alzheimer's Disease Neuroimaging Initiative (ADNI) dataset, allowing our model to learn generalizable features of human brain anatomy. This model is then fine-tuned on a small dataset of vervet brain MRI from The Aging Vervet Colony (AVC) at Wake Forest Alzheimer's Disease Research Center (ADRC) to adapt to the NHP-specific neuroanatomy. This enables accurate segmentation of six key tissue types: grey matter (GM), white matter (WM), CSF, deep grey matter, brainstem, and cerebellum. The combination of STU-Net and transfer learning effectively delineates complex tissue boundaries and captures fine anatomical details specific to NHP brains. Notably, our method demonstrated improvement in segmenting small subcortical structures such as putamen and thalamus that are challenging to resolve with limited spatial resolution and tissue contrast, and achieved DSC of over 0.88, IoU over 0.8 and HD95 under 7. This study introduces a robust method for multi-class brain tissue segmentation in NHPs, potentially accelerating research in evolutionary neuroscience and preclinical studies of neurological disorders relevant to human health.

**Keywords:** Nonhuman primate, brain tissue segmentation, deep learning.

## 1      Introduction

Nonhuman primates (NHPs) serve as valuable models of cognitive aging and various pathological conditions due to their similarities in neuroanatomy, social skills, and psychological characteristics with humans. Normally, NHPs have a lifespan of about 25 years, with developmental stages progressing 4-7 times faster than humans [1], making them ideal for longitudinal studies on brain development and disorders. Thus, besides being particularly useful for studying the impacts of hypertension, reduced gait speed, sarcopenia, and glucose intolerance [2-4], they have been widely used in MRI studies to infer normal human brain development [1, 5-7] as well as developmental disorders [8, 9]. In particular, the NHP model is useful for investigating the effects of drugs and maternal environments on brain structures [8, 10-12]. Among NHPs, vervets are advantageous as Alzheimer's Disease(AD) models due to their close biological proximity to humans, including similarities in brain structure and function, as well as comparable endocrine, social, and cognitive characteristics [13]. Their larger body size also facilitates imaging studies and CSF collection. Notably, vervet Aβ shares 100% sequence identity with human Aβ and forms aggregates in the brain with age. In captivity, vervets can live into their mid-to-late 20's [14] and often exhibit age-related declines in cognitive and physical function [15, 16], Aβ deposition, gliosis, and neuronal dystrophy [17, 18]. Neuritic plaques are also observed, with immunoreactive phosphorylated tau present in dystrophic neurites [18]. However, to accurately interpret structural MRI findings on vervet brain tissue morphology changes and determine disease-specific changes, robust and reliable vervet brain tissue segmentation methods are essential.

Recent years have seen significant advancements in deep learning techniques for imaging analysis of NHP brain. Several studies have leveraged UNet-based models to address NHP brain extraction tasks. HC-Net [19] introduced a hybrid convolutional encoding-decoding structure, combining 2D and 3D convolutions to efficiently extract features from limited macaque MRI data for brain extraction. BEN [20] enhanced the UNet model with non-local attention mechanisms to improve brain extraction performance across species. Wang et al. [21] proposed a transfer learning approach, adapting knowledge learned from human brain MRIs to macaque MRIs for brain extraction.

In contrast, studies focusing specifically on NHP brain tissue segmentation are more limited. nBEST [22] utilized a 3D U-NeXt model with lifelong learning to handle NHP brain MRIs across multiple species, sites and developmental stages, addressing both brain extraction and tissue segmentation. However, to the best of our knowledge, there are no existing models that concentrate specifically on vervet brain tissue segmentation that could provide insights into development of AD.

Despite these advancements, current models still face limitations due to the scarcity of annotated NHP brain MRI datasets [20, 21], the small size of NHP brain [19, 23], limited resolution [19, 23] of available imaging data and the anatomical differences between human and NHP brains [22]. Fig.1 highlights the key anatomical differences between human and vervet brain that impact segmentation accuracy and model performance. As indicated by the red arrows, the human brain exhibits greater cortical folding and deeper sulci, whereas the vervet brain has a smoother cortical surface with less obvious gyrification. Subcortical structures, including caudate, putamen, internal capsules and thalamus, appear less apparent in the NHP brain, making tissue boundaries less distinct compared to the human brain, which contributes to the greater difficulty for NHP brain tissue segmentation than humans. Additionally, due to the small size of the vervet brain, the resolution of NHP brain MRI is limited compared to the human brain MRI.

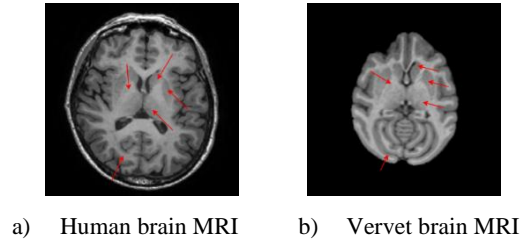

a)   Human brain MRI     b)   Vervet brain MRI

**Fig. 1.** Human and vervet brain MRI comparison. a) sample axial slice of human brain MRI; b) sample axial slice of vervet brain MRI. (Red arrows indicate four sub-cortical structure: caudate, putamen, internal capsules and thalamus, and White Matter and Gray Matter tracts around the boundaries).

To address these limitations and fill the existing research gap, we decided to take advantage of a valuable vervet brain MRI dataset, The Aging Vervet Colony (AVC) at Wake Forest Alzheimer's Disease Research Center (ADRC), and propose a novel approach for vervet brain tissue segmentation using transfer learning. Our model, based on the STU-Net architecture, aims to leverage the sufficient human brain MRI data to overcome the scarcity of annotated vervet brain MRI images via transfer learning. By pretraining the model on a large human brain MRI dataset (ADNI) and subsequently fine-tuning it on a much smaller vervet brain MRI dataset (a small subset of AVC dataset), we seek to develop a robust and accurate segmentation tool specifically tailored for vervet brains to accurately segment the six key brain tissues: Cerebrospinal Fluid (CSF), Gray Matter (GM), White Matter (WM), Deep Gray Matter (DGM), brainstem and cerebellum. This approach not only addresses the challenge of limited NHP data but also enables the exploration of potential links between vervet brain structure and AD development, opening new avenues for comparative neuroscience research.

## 2   Materials and Methods

In this study, we employ STU-Net as the backbone and incorporate transfer learning to develop a specialized model for vervet brain tissue segmentation. Our approach involves two key stages: (1) pretraining the model on ADNI, a large human brain MRI dataset, for human brain tissue segmentation to learn the anatomical features of the brain, and (2) fine-tuning the pretrained model on a small subset of AVC, a curated vervet brain MRI dataset to transfer the knowledge learnt from human brain to vervet brain. We evaluate the model's performance both quantitatively and qualitatively. Quantitative assessment includes comparison of metrics such as Dice score, Intersection over Union (IoU), and 95th percentile Hausdorff Distance. Qualitative evaluation involves visual inspection of segmentation maps, comparing our method against the traditional ANTs algorithm and two benchmark deep learning models for medical image segmentation: nnUNetV2 basic and nnUNet ResEnc L. This comprehensive evaluation allows us to assess the effectiveness of our transfer learning approach in addressing the challenges of vervet brain tissue segmentation.

### 2.1  ANTs

ANTs [24] is a widely used medical image registration and segmentation toolkit. Developed as an open-source software library, ANTs offers a comprehensive set of tools for processing and analyzing medical images. It incorporates N4 bias field correction and Atropos for brain MRI tissue segmentation. Atropos is a multivariate n-class segmentation algorithm using a generalized Expectation Maximization (EM) algorithm to classify brain tissues. It incorporates prior information through Markov Random Fields (MRFs) and template-based spatial priors, and can handle multivariate image data and multiple classes.

### 2.2  nnUNet and STU-Net

The nnUNet (no-new-UNet)[25] framework, introduced by Isensee et al., is a widely recognized model for medical image segmentation that automatically adapts to the given segmentation task, optimizing preprocessing, network architecture, training, and post-processing. Building upon the foundational UNet architecture with its characteristic encoding-decoding structure, nnUNet enhances the original design by incorporating self-adapting features, including dynamic network configuration, 2D and 3D model combinations and advanced training techniques, enabling it to consistently outperform tradition UNet models across diverse medical image segmentation tasks.

As an extension of nnUNet, STU-Net(Scalable and Transferable U-Net) [26] was developed by Huang et al. to tackle the scalability limitations of current medical image segmentation models. It offers a series of models with parameter sizes ranging from 14 million to 1.4 billion. Being pre-trained on a large-scale TotalSegmentator dataset, STU-Net demonstrates strong transferability across ultiple downstream tasks. Thus, in our study, we leverage STU-Net's adaptability to handle the unique challenges presented by vervet brain MRI data, particularly its ability to optimize performance with limited training samples.

### 2.3  Transfer learning

Transfer learning is a deep learning technique that utilizes knowledge gained from solving one problem to address a related but different problem. In the context of our study, we employ transfer learning to overcome the limitations of scarce annotated vervet brain MRI data by taking advantage of the knowledge learnt from human brain MRI data. Our approach involves:

1. Pretraining: The nnUNet model is initially trained on ADNI dataset with sufficient human brain MRI data;
2. Fine-tuning: The pretrained model is then adapted to the vervet brain domain using the limited available vervet MRI data from AVC dataset at Wake Forest ADRC

This transfer learning strategy allows us to leverage the features learned from human brain MRIs while adapting to the specific characteristics of vervet brain anatomy.

### 2.4    Dataset

Our study utilizes two primary datasets, ADNI and AVC, for pre-training and fine-tuning, respectively.

The Alzheimer's Disease Neuroimaging Initiative (ADNI) dataset is a comprehensive collection of longitudinal clinical, imaging, genetic, and biomarker data widely used in AD research. ADNI uses 3T scanners from three major vendors: GE, Philips, and Siemens. The imaging protocol for T1-weighted scans uses an MP-RAGE sequence with a resolution of 1x1x1mm³ and a field of view of 208x240x256mm. To create our pre-training dataset, we selected 100 cases from ADNI dataset and use the manual segmentation by radiologists as groundtruth for training.

The Aging Vervet Colony (AVC) at the Wake Forest Alzheimer's Disease Research Center (ADRC) has collected brain magnetic resonance imaging (MRI), blood and CSF samples, and cognitive function assessments for the past four years from middle-aged and elderly vervets, ranging in age from 8 years to end of life (~29 years), providing a comprehensive longitudinal dataset. T1-weighted anatomic images were acquired using a 3T Siemens Skyra scanner. The imaging protocol employed MPRAGE sequence, with voxel dimensions of 0.5×0.5×0.5 mm³. For the model fine-tuning, we selected 12 cases for training and 6 cases for validation and used the manual segmentations from radiologist as groundtruth.

### 2.5    Training details

The training data consists of brain-only images. The vervet brain MRI has been resampled to match the resolution of the human brain MRI. Throughout both the training and inference processes, each image was clipped at the 99.9th percentile to mitigate the effects of outlier intensities and then normalized using Min-Max Scaling individually. All models are implemented using Pytorch library. The pre-trained human model was trained for 1,000 epochs with an initial learning rate of 1e-2, followed by fine-tuning on the monkey dataset for 200 epochs with an initial learning rate of 1e-4. Various data augmentation techniques were employed during training, including rotations, scaling, Gaussian noise, Gaussian blur, brightness and contrast adjustments, low-resolution simulation, gamma transformations, and mirroring. These augmentations increase the diversity of the training data, thereby enhancing the network's generalization performance. All experiments were conducted on NVIDIA RTX A6000 GPUs with 48GB VRAM.

## 3   Results

### 3.1   Evaluation Metrics

To evaluate the performance of our model on vervet brain tissue segmentation, several evaluation metrics have been employed, including Dice Similarity Coefficient (DSC), Intersection over Union (IoU) and 95th percentile Hausdorff Distance (HD95). Both DSC and IoU measure the overlap between the predicted segmentation and the groundtruth, ranging from 0 to 1, with 1 indicating perfect overlap. HD95 measures the maximum distance between the boundaries of the predicted segmentation and groundtruth while discarding 5% outlier values to mitigate the sensitivity to outliers in the standard Hausdorff distance.

### 3.2   Comparison with other methods

As illustrated in Table 1, Our model outperforms the traditional ANTs method and two other deep learning medical image segmentation benchmarks nnUNetV2 basic and nnUNet ResEnc L in terms of overall Dice Score, IoU and 95th percentile Hausdorff Distance across all six brain tissue types. While our model slightly underperforms nnUNetV2 in sub-cortical specific areas, it still demonstrates comparable and very close performance and outperforms the other two models in these regions. Notably, our model excels in boundary delineation, as evidenced by the significantly decreased 95th percentile Hausdorff Distance. It reduces the distance by half compared to the traditional ANTs method in overall evaluation and shows significant improvements over the other two deep learning benchmarks across all six tissue types. In sub-cortical specific areas, our model still outperforms all three comparative methods, further highlighting its effectiveness in outlining complex brain structures. These results underscore the robustness and accuracy of our transfer learning approach in vervet brain tissue segmentation, particularly in capturing fine anatomical details and boundaries.

**Table 1.** Evaluation metrics comparison between models. (DSC - Dice Score; IoU - Intersection over Union; HD95 - 95th Percentile Hausdorff Distance; Overall – average over 6 types of brain tissues; Sub-cortical – only calculated based on sub-cortical segmentations)

| Model | DSC | | IoU | | HD95 | |
|---|---|---|---|---|---|---|
| | Overall | Sub-cortical | Overall | Sub-cortical | Overall | Sub-cortical |
| nnUNetV2 basic | 0.8727 | 0.8833 | 0.7904 | 0.7919 | 8.7832 | 7.2816 |
| **Our Method** | **0.8843** | 0.8891 | **0.8059** | 0.8013 | **6.9927** | **6.9507** |
| nnUNet ResEnc L | 0.8828 | **0.8893** | 0.8033 | **0.8019** | 8.4396 | 7.0321 |
| ANTs | 0.7998 | 0.7638 | 0.6889 | 0.6182 | 14.1072 | 7.7786 |

Fig.2 provides a qualitative comparison of the segmentation among the three deep learning models and the traditional ANTs method. Overall, the deep learning models demonstrate impressive performance in delineating the boundaries of structures such as the putamen (as indicated by the red arrows in Fig. 2d)). However, they exhibit some discontinuity in the thalamus region compared to the traditional ANTs method (as indicated by the red arrows in Fig. 2a) to Fig. 2c)). Among the three deep learning models, our model displays the best performance, with the least missing parts in the thalamus areas (as indicated by the red arrows in Fig. 2a) to Fig. 2c)). This visual assessment aligns with our model's low HD95. By comparing the segmentation results from Fig. 2a) to Fig. 2d) with the ground truth in Fig. 2e), it is evident that our method achieves the closest delineation of structural boundaries to groundtruth, particularly in complex regions such as the thalamus.

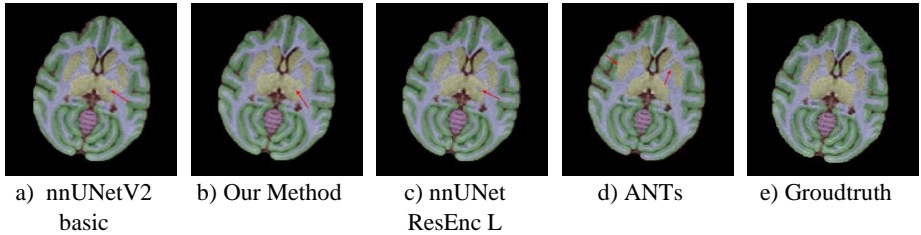

a) nnUNetV2 basic    b) Our Method    c) nnUNet ResEnc L    d) ANTs    e) Groudtruth

**Fig. 2.** Comparison of segmentation details between models a) segmentation map generated by nnUNetV2 basic; b) segmentation map generated by our method; c) segmentation map generated by nnUNet ResEnc L; d) segmentation map generated by traditional ANTs method; e) manual segmentation as groudtruth.

These qualitative observations corroborate our quantitative findings, highlighting our model's ability to capture fine anatomical details and maintain structural continuity. The improved boundary delineation and reduced discontinuities in sub-cortical brain structures emphasize the effectiveness of our transfer learning approach in adapting to the specific challenges of vervet brain tissue segmentation.

To statistically compare the performance of the four methods, a Kruskal-Wallis test was conducted across them (Our method, nnUNet ResEnc L, nnUNetV2 basic, and ANTs) using three different evaluation metrics (DSC, IoU, and HD95). The test revealed statistically significant differences among the models in all three experiments (H-statistic = 30.1871, 30.3427, and 24.7594; $p < 0.0001$), indicating that at least one model performed significantly differently from the others. To determine pairwise differences, Dunn's post-hoc test was performed. The results consistently showed that ANTs significantly differed from the other methods (p-values $< 0.05$ in multiple comparisons) across all three evaluation metrics. In contrast, our method, nnUNet ResEnc L, and nnUNetV2 basic generally did not show significant differences among themselves (p-values $> 0.05$), suggesting comparable performance.

These findings indicate that ANTs consistently differs significantly from the other three methods, while the performance of the remaining UNet based deep learning methods is relatively similar across different evaluation metrics.

## 4      Discussion

In this study, we aimed to develop and evaluate a transfer learning-based approach for non-human primate (NHP) brain tissue segmentation. Given the scarcity of large, annotated NHP neuroimaging datasets, we leveraged transfer learning to adapt models trained on human MRI data for application in NHP brain segmentation. Our results demonstrate that this approach effectively segments six key brain tissue structures such as basal ganglia regions with high accuracy. Our model has been quantitatively and qualitatively compared with the conventional method ANTs and two other benchmarks for medical image segmentation for evaluation. Compared to conventional methods such as ANTs, our model achieved substantial improvements in segmentation performance delineating putamen and thalamus boundaries and significantly improved the metrics including Dice Score, IoU and HD95, highlighting the contribution of deep learning and transfer learning in this task. Although there is no statistically significant difference among the three U-Net based deep learning models including ours, our method still outperform the other two regarding most of the evaluation metrics and visual inspection. Specifically, our model achieved the best performance regarding the overall segmentation across six different tissue types with respect to all three metrics. Although slightly underperforms nnUNet ResEnc L in the sub-cortical specific regions, our model shows very close performance and displayed impressive performance in terms of HD95, meaning that it can best capture the boundaries of the structure. Thus, the ability of our model to generalize with minimal training data demonstrates its potential for broader applications.

This study has significant implications for AD research, particularly in translational studies that utilize NHP models to investigate neurodegeneration. The ability to accurately segment NHP brain tissue enhances the precision of structural and functional analyses, facilitating cross-species comparisons with human AD pathology. Given the increasing interest in using NHP models for preclinical AD studies, our approach offers a valuable tool for quantitative neuroimaging assessments such as cortical thickness and brain tissue volume, potentially improving biomarker identification and treatment evaluation in translational neuroscience[27-30].

Despite its promising results, this study has several limitations. First, the training dataset was relatively small, which may limit the model's ability to generalize to diverse NHP populations. Second, the study was conducted on a single NHP species, which may restrict its applicability to other species with different brain morphologies. Third, the reliance on T1-weighted MRI data alone may not fully capture tissue contrasts relevant for precise segmentation. Future studies should address these limitations by incorporating larger, more diverse datasets and multimodal MRI sequences.

In conclusion, we have demonstrated the feasibility of transfer learning for NHP brain tissue segmentation, achieving robust performance with limited training data. Moving forward, we propose the integration of T2-weighted and diffusion MRI to enhance tissue contrast and segmentation accuracy. Additionally, expanding the approach to multiple NHP species could improve model generalizability across different anatomical structures. Furthermore, incorporating state-of-the-art techniques in latest deep learning models may further refine segmentation accuracy and applicability in translational neuroimaging studies.